\documentclass[manuscript]{acmart}

\usepackage{soul}
\usepackage{multirow}
\usepackage{fancybox}
\usepackage{enumitem}
\usepackage{graphicx}
\usepackage{balance}
\usepackage{color}
\usepackage{float}
\usepackage{adjustbox}
\usepackage{xcolor}
\usepackage{array}
\usepackage{amsmath}
\usepackage{xspace}
\usepackage{url}
\usepackage{tikz}
\usepackage{caption}
\usepackage{pgfplots}
\usepackage{listings}
\usepackage{color}
\usepackage{graphicx}
\definecolor{dkgreen}{rgb}{0,0.6,0}
\definecolor{gray}{rgb}{0.5,0.5,0.5}
\definecolor{mauve}{rgb}{0.58,0,0.82}
\pgfplotsset{compat=1.16}
\usepackage{pgf-pie}
\usetikzlibrary{pgfplots.statistics,calc}
\usepackage{algorithm}
\usepackage{algpseudocode}
\usepackage{subcaption}
\usepackage{listings}
\usepackage{tabularx}

\usepackage{tcolorbox}

\makeatletter
\newcommand{\mybox}[1]{%
	\setbox0=\hbox{#1}%
	\setlength{\@tempdima}{\dimexpr\wd0+13pt}%
	\begin{tcolorbox}[boxrule=0.5pt, colback=white, arc=4pt,
		left=6pt,right=6pt,top=6pt,bottom=6pt,boxsep=0pt]
		#1
	\end{tcolorbox}
}

\definecolor{codegreen}{rgb}{0,0.6,0}
\definecolor{codegray}{rgb}{0.5,0.5,0.5}
\definecolor{codepurple}{rgb}{0.58,0,0.82}
\definecolor{backcolour}{rgb}{0.95,0.95,0.92}

\lstdefinestyle{mystyle}{
  language=Python,
  aboveskip=3mm,
  showstringspaces=false,
  columns=flexible,
  numbers=none,
  backgroundcolor=\color{backcolour},
  commentstyle=\color{codegreen},
 keywordstyle=\color{magenta},
    numberstyle=\tiny\color{codegray},
    stringstyle=\color{codepurple},
    basicstyle=\small\ttfamily,
    breakatwhitespace=false,         
    breaklines=false,                 
    captionpos=b,                    
    keepspaces=false,                 
    numbersep=5pt,                  
    showspaces=false,                
    showstringspaces=false,
    showtabs=false,                  
    tabsize=2,
    escapeinside=``
}
\definecolor{nima2}{RGB}{1.0, 0.49, 0.0}
\definecolor{songcolor}{RGB}{191,191,191}
\definecolor{nimacolor}{RGB}{0.13, 0.67, 0.8}
\definecolor{aruncolor}{RGB}{51,255,51}

\AtBeginDocument{%
  \providecommand\BibTeX{{%
    \normalfont B\kern-0.5em{\scshape i\kern-0.25em b}\kern-0.8em\TeX}}}

\setcopyright{acmcopyright}
\copyrightyear{2018}
\acmYear{2018}
\acmDOI{XXXXXXX.XXXXXXX}

\acmBooktitle{Woodstock'18: ACM Symposium on Neural Gaze Detection,
 June 03--05, 2018, Woodstock, NY} 
\acmPrice{15.00}
\acmISBN{978-1-4503-XXXX-X/18/06}

\begin{document}
\newpage


\title{A Pilot Study on LLM-Based Agentic Translation from Android to iOS: Pitfalls and Insights}

 \author{Zhili Zeng} 
 \authornote{Both authors contributed equally to this research.}
 \email{z75zeng@yorku.ca}
 \affiliation{%
   \institution{York University}
   \streetaddress{4700 Keele St.}
   \city{North York}
   \state{Ontario}
   \country{Canada}
   \postcode{M3J 1P3}
 }

\author{Kimya Khakzad Shahandashti} 
\authornotemark[1]
 \email{kimya@yorku.ca}
 \affiliation{%
   \institution{York University}
   \streetaddress{4700 Keele St.}
   \city{North York}
   \state{Ontario}
   \country{Canada}
   \postcode{M3J 1P3}
 }

 \author{Alvine Boaye Belle}
 \email{alvine.belle@lassonde.yorku.ca}
 \affiliation{%
   \institution{York University}
   \streetaddress{4700 Keele St.}
   \city{North York}
   \state{Ontario}
   \country{Canada}
   \postcode{M3J 1P3}
 }
 \author{Song Wang}
 \email{wangsong@yorku.ca}
 \affiliation{%
   \institution{York University}
   \streetaddress{4700 Keele St.}
   \city{North York}
   \state{Ontario}
   \country{Canada}
   \postcode{M3J 1P3}
 }
  \author{Zhen Ming (Jack) Jiang}
 \email{zmjiang@eecs.yorku.ca}
 \affiliation{%
   \institution{York University}
   \streetaddress{4700 Keele St.}
   \city{North York}
   \state{Ontario}
   \country{Canada}
   \postcode{M3J 1P3}
 }

\begin{abstract}
The rapid advancement of mobile applications has led to a significant demand for cross-platform compatibility, particularly between the Android and iOS platforms. Traditional approaches to mobile application translation often rely on manual intervention or rule-based systems, which are labor-intensive and time-consuming. While recent advancements in machine learning have introduced automated methods, they often lack contextual understanding and adaptability, resulting in suboptimal translations.  

Large Language Models (LLMs) were recently leveraged to enhance code translation at different granularities, including the method, class, and repository levels. Researchers have investigated common errors, limitations, and potential strategies to improve these tasks. However, LLM-based application translation across different platforms, such as migrating mobile applications between Android and iOS or adapting software across diverse frameworks, remains underexplored. 
Understanding the performance, strengths, and limitations of LLMs in cross-platform application translation is critical for advancing software engineering automation. This study aims to fill this gap by evaluating LLM-based agentic approaches for mobile application translation, identifying key failure points, and proposing guidelines to improve translation performance.  

Specifically, we experimented with five Android projects of varying sizes, ranging from small (fewer than 3K LOC) to large (over 149K LOC), to ensure a diverse evaluation across different codebases. 
We developed a chain of agents that account for dependencies, specifications, program structure, and program control flow when translating applications from Android to iOS. 
To evaluate the performance, we manually examined the translated code for syntactic correctness, semantic accuracy, and functional completeness. For translation failures or partial results, we further conducted a detailed root cause analysis to understand the underlying limitations of the agentic translation process and identify opportunities for improvement. 
Our findings aim to shed light on the readiness of LLMs for mobile cross-platform development, identifying key challenges and opportunities for improving the performance of this complex translation process. 
\end{abstract}



\maketitle

\section{Introduction}
\label{sec:introduction}


Code translation plays a critical role in modern software engineering, enabling developers to reuse functionality, adapt to new platforms, and maintain consistency across diverse ecosystems~\cite{nguyen2014migrating}. As mobile applications proliferate across multiple operating systems, translating apps between platforms such as Android and iOS has become increasingly critical for maintaining consistent user experiences and reducing redundant development efforts across platforms~\cite{lazareska2017analysis}. 



Recently, the emergence of large language models (LLMs) has opened new possibilities for automating code translation. By leveraging vast training corpora and deep contextual understanding, LLMs such as GPT-4o~\cite{openai2023gpt4} have shown promising capabilities in generating and translating code across programming languages. These models have achieved notable success in method-level and class-level translation tasks, such as converting code snippets or classes between Java and Python~\cite{bairi2024codeplan,yuan2024transagent,yan2023codetransocean,pan2024lost,jain2023codeplan,saha2024specification,ibrahimzada2024alphatrans}. However, despite these advancements, the performance and applicability of LLM-based translation approaches in the context of mobile application migration, especially across fundamentally different platforms like Android and iOS, remain largely unexplored. Mobile apps present unique challenges, including platform-specific APIs, lifecycle management, UI paradigms, and tightly integrated architectural patterns. Existing studies~\cite{bairi2024codeplan,yuan2024transagent,yan2023codetransocean,pan2024lost,jain2023codeplan,saha2024specification,ibrahimzada2024alphatrans} typically evaluate LLMs on narrowly scoped (i.e., method-level or class-level), syntactic transformations, offering limited insight into how these models perform when tasked with end-to-end translation of real-world mobile applications.

In this work, we investigated the capabilities and limitations of LLM-based agentic approaches for mobile application translation, taking Android-to-iOS migration as a representative and practically significant case study. We designed and conducted an experimental study that systematically assesses how well state-of-the-art LLMs can translate open-source Android applications into functionally equivalent iOS applications. 

Specifically, we conducted experiments using five open-source Android projects of varying sizes, ranging from small applications with fewer than 3K lines of code (LOC) to larger projects exceeding 149K LOC. 
This selection ensures a diverse evaluation across different application domains and complexities. To perform the translation from Android to iOS, we developed a chain of specialized agents designed to consider key aspects such as library dependencies, high-level specifications, program structure, and control flow. 
To evaluate the effectiveness of our approach, we manually analyzed the translated code with respect to syntactic correctness, semantic accuracy, and functional completeness. For translation failures or partial results, we further conducted a detailed root cause analysis to understand the underlying limitations of the LLM-based agentic translation process and identify opportunities for improvement. 
Based on the findings from our experiments, we further  identified five key challenges and opportunities in improving the performance of this complex translation process, which aims to shed light on the readiness of LLMs for mobile cross-platform development.  The contributions of this paper are: 

\begin{itemize}
\item We designed and implemented a multi-agent translation pipeline that integrates dependency resolution, specification interpretation, and control-flow analysis to guide LLMs in performing accurate and context-aware cross-platform translation.

\item We presented the first systematic experimental study evaluating the effectiveness of a state-of-the-art LLM-based agentic approach for translating real-world Android applications to iOS, covering projects of varying sizes and application domains.

\item We conducted a detailed manual analysis of the translated iOS code to assess both syntactic correctness and semantic accuracy, and we performed a root cause analysis for failed or partially correct translations.

\item We identified key limitations of current LLM-based translation approaches and provided practical guidelines and future research directions to improve agentic mobile application translation.

\item We publicly released the data and code from our study to facilitate reproducibility and support future research and development in LLM-based mobile application translation\footnote{\url{https://zenodo.org/records/15881864}}.
\end{itemize}

The remainder of this paper is organized as follows: Section~\ref{sec:2} introduces the background of this work, 
Section~\ref{sec:3} presents our multi-agent translation pipeline for Android-to-iOS translation, 
Section~\ref{sec:4} shows our experiment settings,
Section~\ref{sec:5} discusses the results, 
Section~\ref{sec:6} discusses the threats to validity of this work, and Section~\ref{sec:7} concludes this work. 
\section{Background and Related Work}
\label{sec:2}

This section provides an overview of code translation methods and the emerging role of LLMs in software engineering tasks.

\subsection{Code Translation}
Code translation involves converting source code from one programming language to another while maintaining the original program's functionality \cite{xia2023empirical}. Traditional methods of code translation \cite{C2Rust, Sharpen, phan2017statistical, pawlak2016spoon} have largely depended on rule-based systems and syntax-directed translation schemes \cite{hong2023improving}. However, these methods often face challenges when dealing with languages with different paradigms or lacking a one-to-one correspondence in features, which can result in semantic discrepancies and bugs in the translated code \cite{wang2023lost}. Recent advancements have introduced machine learning techniques to enhance code translation. For example, \textit{CodeTransOcean} offers a multilingual benchmark for code translation, aiding in the training and evaluation of models across various programming languages \cite{zhong2023codetransocean}. Nevertheless, many of these approaches tend to focus on translating code at the function or snippet level, which does not adequately address the complexities found in real-world code repositories. These complexities can include intricate dependencies, diverse coding styles, and various architectural patterns \cite{leclair2023translating}. Additionally, research has indicated that LLMs can inadvertently introduce subtle bugs during code translation due to misunderstandings of the context or language-specific nuances \cite{wang2023lost}. This underscores the necessity for more advanced methods that can perform repository-level translations while minimizing the risk of errors. 
Jana et al.~\cite{jana2024cotran} fine-tune the LLM via compiler and symbolic-based testing feedback to implement multiple programming language translations. Pan et al.~\cite{pan2024lost} manually investigate the reasons for unsuccessful translations and create a taxonomy of 15 translation bugs. 
Bairi et al.~\cite{bairi2024codeplan} introduced CodePlan, an agent for repository-level coding tasks. CodePlan synthesized a multi-step chain-of-edits (plan), where each step results in a call to an LLM on a code location with context derived from the entire repository, previous code changes, and task-specific instructions. 
Ibrahimzada et al.~\cite{ibrahimzada2024alphatrans} proposed AlphaTrans, a neuro-symbolic approach to automate repository-level code translation. AlphaTrans translates both source and test code, and employs multiple levels of validation to ensure the translation preserves the functionality of the source program. 

However, despite these advancements, the performance and applicability of LLM-based translation approaches in the context of mobile application migration, especially across fundamentally different platforms like Android and iOS, remain largely unexplored. In this work, we investigate the capabilities and limitations of LLM-based approaches for agentic mobile application translation, taking Android-to-iOS migration as a representative and practically significant case study.  

\subsection{LLMs in Software Engineering}
Large Language Models (LLMs) such as GPT-4 \cite{openai2023gpt4} and Codex \cite{chen2021evaluating}, have demonstrated remarkable capabilities in natural language understanding and code generation. In software engineering, LLMs have been applied to various tasks, including code generation~\cite{svyatkovskiy2020intellicode,mu2024clarifygpt}, bug detection \cite{allamanis2018survey}, code summarization~\cite{ahmad2020transformer,shin2023prompt}, test generation~\cite{shin2024retrieval,chen2024chatunitest}, code translation~\cite{saha2024specification,jana2024cotran,pan2024lost,bairi2024codeplan}, and automated documentation \cite{roy2021reassessing}. LLMs leverage deep learning architectures, particularly transformer models \cite{vaswani2017attention}, to capture long-range dependencies and contextual information within code. This makes them well-suited for tasks that require understanding complex code structures and semantics. In the context of code translation, LLMs have shown potential in translating code snippets between languages \cite{roziere2020unsupervised}. Techniques like retrieval-augmented generation have been proposed to enhance the few-shot learning capabilities of LLMs for code translation tasks \cite{he2023enhancing}. Additionally, systems like \textit{CodePlan} \cite{jain2023codeplan} and \textit{TRANSAGENT} \cite{liang2023transagent} have begun to explore repository-level code translation by incorporating planning and multi-agent strategies to handle the complexities of larger codebases. Despite these advancements, challenges remain in scaling LLMs to handle repository-level translation effectively. Issues such as limited context windows, managing inter-file dependencies, and ensuring the functional correctness of translated code require further research \cite{jain2023codeplan, leclair2023translating}.

\subsection{Agent in LLMs}

LLM agents are artificial entities that leverage LLMs to automate the reasoning, planning, and responding actions in specific surrounding environments~\cite{xi2025rise, wang2024survey}. LLM-based agents apply in multiple software engineering areas \cite{liu2024large}. Specifically, LLM-based agents' exploration in the fields of mobile development \cite{zhang2023appagent, rawles2023androidinthewild} has attracted significant attention, and has led to a variety of practices and the emergence of benchmarks \cite{wang2024mobileagentbench}. A basic LLM-based agent architecture consists of planning, memory, perception, and action procedure \cite{liu2024large}. Particularly, a) the planning procedure divides complex tasks into smaller sub-tasks and re-organizes them to accomplish the final goals; b) the memory procedure acquires metadata from the environment and utilizes them to support future actions \cite{wang2024survey}; c) the perception procedure further processes stored information (i.e. textual, visual or auditory data) with customized mechanism \cite{qian2023communicative}; d) and the action procedure interacts with the environment by translating the agent’s decisions into specific outcomes. The agents proposed in our study undergo adjustment and derivation of this basic structure, and we aggregate multiple components into an automated pipeline through the coupling of inputs and outputs across agents.

\section{Approach} 
\label{sec:3}

\begin{figure}[t!] \centering \includegraphics[width=0.8\linewidth]{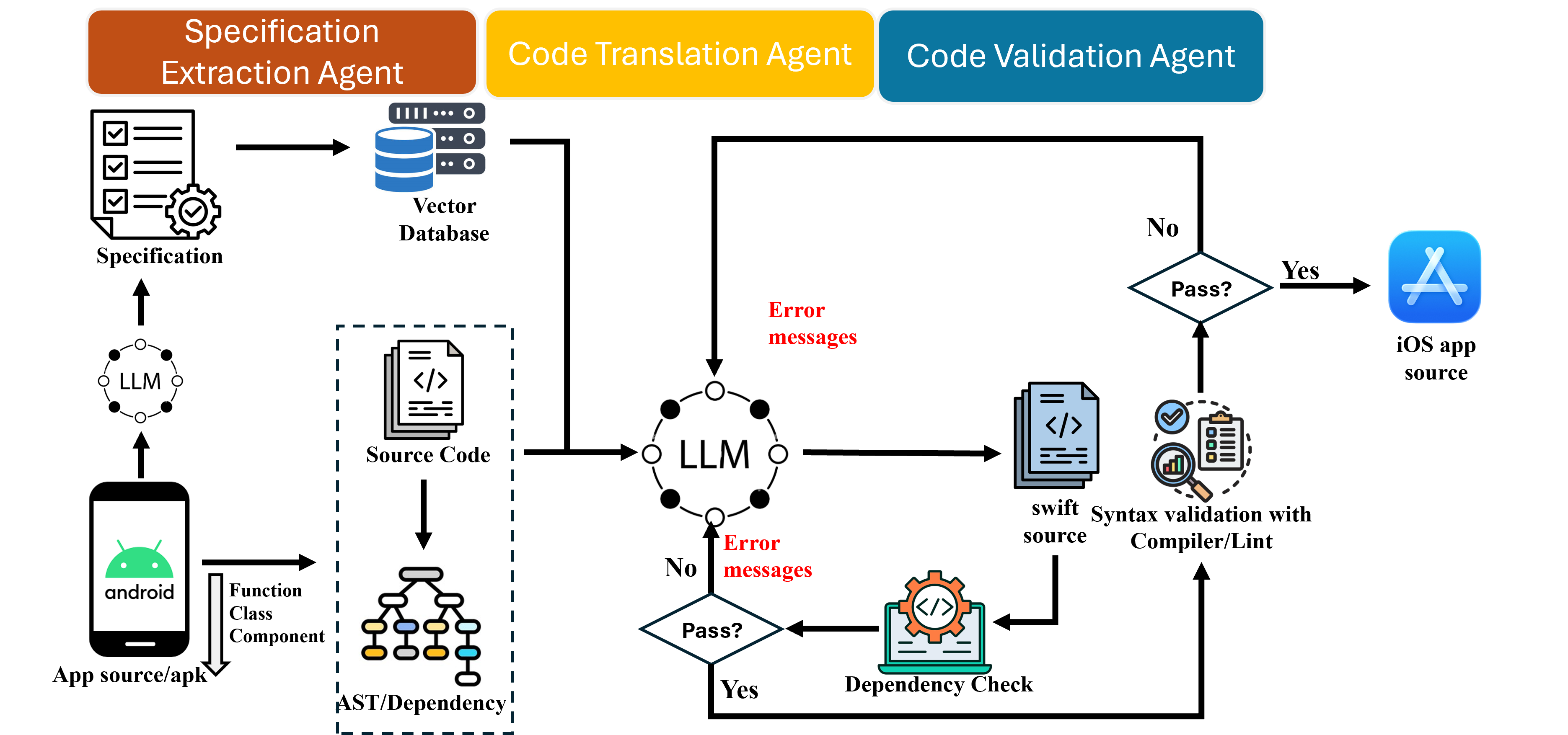} \caption{Overview of our LLM-based agent for Android-to-iOS translation} \label{fig:code_overview} \end{figure}
Our proposed approach leverages LLMs for Android-to-iOS translation through a coordinated, \emph{agent-based architecture} that incorporates \emph{context-aware translation} and \emph{validation} strategies. By modularizing tasks into specialized agents, we address the challenges of translating complex, real-world codebases systematically and robustly. 
Specifically, the proposed approach is built around three specialized agents, each designed to handle a distinct phase of the code translation workflow: (1) the Specification Extraction Agent, (2) the Code Translation Agent, and (3) the Code Validation Agent. 
These agents operate in a coordinated fashion to improve translation accuracy, maintain functional integrity, and support scalable automation across diverse codebases. 
Figure~\ref{fig:code_overview} highlights the key phases of our approach. The detailed workflow of each agent is as follows: 

\begin{enumerate} 

\item \textbf{Specification Extraction Agent}: This agent leverages Retrieval-Augmented Generation (RAG)~\cite{lewis2020retrieval} to automatically retrieve, index, and synthesize relevant information from project documentation, code comments, pull requests, issues, and extracted specifications for the project's website. The extracted specifications help provide contextual grounding for the translation task, ensuring that platform-specific behaviors and requirements are accurately captured and preserved in the translated output. 

\item \textbf{Code Translation Agent}: This agent performs comprehensive static analysis, such as Abstract Syntax Tree (AST) parsing and dependency graph construction, to extract structural and semantic context from the source code. Leveraging this information along with extracted specifications, the agent constructs rich, context-aware prompts to guide the LLM in translating Android-specific functions and components into their iOS equivalents. The translated code is then integrated into the appropriate class and module structures to preserve application architecture, maintain behavioral consistency, and align with platform conventions. 

\item \textbf{Code Validation Agent}: This agent ensures the correctness and integrity of the translated code through multiple validation stages. The agent first  
verifies semantic consistency by analyzing inter-file references and class dependencies. It ensures that module imports, platform-specific calls, and cross-component interactions remain functional and coherent in the translated iOS codebase, reducing runtime failures and integration issues. It then performs automated linting and syntax checking to detect and resolve language-specific errors. 

\end{enumerate}

These agents work collaboratively to translate a given Android application into its iOS equivalent. The Specification Extraction Agent runs only once, while the Code Translation Agent iteratively improves translations based on feedback from the Code Validation Agent.

\subsection{Specification Extraction Agent}
\label{sec:3.1}

This agent gathers and organizes contextual information to provide comprehensive support for accurate code translation. Leveraging \textbf{Retrieval-Augmented Generation (RAG)}~\cite{lewis2020retrieval}, the agent builds a robust knowledge base for each project based on the project documentation, code comments,
pull requests, issues, and possible external documents in the project's official website. This knowledge base is stored in a vector database and utilized during translation to enhance the LLM’s understanding of the codebase. This agent performs the following tasks:
\begin{itemize}
    \item \textbf{File Processing}: Identifies and reads all relevant text-based files in the repository, such as README files, API documents, pull requests, and issues.
    \item \textbf{Web Crawling}: Crawls the project’s official website and related documentation for additional context.
    \item \textbf{Text Embedding}: 
   Converts processed text into vector embeddings using \texttt{SentenceTransformer}~\cite{reimers-2019-sentence-bert}, a pre-trained model architecture built on top of BERT and its variants. SentenceTransformer is optimized explicitly for generating semantically meaningful sentence-level embeddings, enabling efficient similarity comparison, clustering, and retrieval.  
    \item \textbf{Vector Database Creation}: Uses \texttt{ChromaDB}\footnote{https://www.trychroma.com/}, a persistent vector database library, to store embeddings alongside metadata, such as file paths or URLs, for future retrieval. ChromaDB boasts several key features that enhance its functionality. Firstly, it offers efficient contextual retrieval, ensuring that relevant information is quickly and accurately accessed, which reduces overhead in downstream processes. Additionally, its adaptability allows the agent to be extended to support other data sources or embedding models, making it a versatile choice for various projects.
    \item \textbf{Query Handling}: Allows downstream processes to query the database for relevant information, returning the most contextually similar documents to support translation. For example, given a method from a class to be translated, the system can retrieve associated documentation, usage examples, related API references, or similar method implementations from the database. This contextual retrieval helps the translation agent better understand the intent and expected behavior of the source method, enabling it to generate more accurate, idiomatic, and platform-aligned target code. Additionally, by surfacing relevant dependencies and patterns, it reduces ambiguity and enhances consistency across the translated application. 
\end{itemize}

\subsection{Code Translation Agent}
The Code Translation Agent provides structural and semantic insights into the Android codebase to enable accurate and context-aware translation to iOS. It preserves key code relationships and interdependencies critical for maintaining the original application's functionality and structure. 
To achieve this, the agent leverages Tree-sitter~\cite{treesitter} to parse source files into Abstract Syntax Trees (ASTs), which capture the syntactic and hierarchical structure of the code. From these ASTs, the agent extracts detailed dependency information at the method, class, and component levels. This includes identifying method calls, class inheritance, import relationships, and module/package-level couplings. The translation follows a dependency-aware, bottom-up strategy to ensure that code elements with fewer external dependencies are translated first, reducing the risk of cascading translation errors. The process consists of the following steps:
\begin{itemize}

\item \textbf{Translation plan}: The agent identifies the component (e.g., module or package) with the fewest dependencies on other components and selects it as the translation starting point. Within the selected component, classes are ranked by their dependency degree. The class with the fewest dependencies on other classes is chosen first. Within the selected class, methods are analyzed and ordered based on their internal dependencies. Methods with the fewest calls to other methods are translated first to provide foundational building blocks.

\item \textbf{Integration}: Once all methods of a class are translated, the agent reconstructs the class structure, integrating the translated methods while preserving inheritance and interface contracts.

\item \textbf{Iterative progression}: After completing one class, the agent proceeds to the next class in descending order of dependency, repeating the translation and integration steps. Once all classes in the current component are processed, the agent moves to the next component, again based on the descending order of inter-component dependencies.

\end{itemize}

For method-level translation, the agent receives method-specific prompts generated from our static analyzer output. These prompts include the method’s AST, its corresponding Java source code, and specification information retrieved from the Specification Extraction Agent (see Section~\ref{sec:3.1}), which describes the method’s usage, intent, and constraints. Using this rich contextual information, the agent employs an LLM to translate Java methods into their Swift equivalents, to preserve the original semantics. Focusing on the method level streamlines the translation process by reducing structural complexity, enabling the agent to work with smaller, self-contained code units. The method-level translator performs the following steps:

\begin{enumerate}
    \item \textbf{AST Analysis}: Leverages output from our static analyzer to identify key structural elements of the code. This includes extracting method-level metadata such as method names, start and end positions within the source file, and associated AST slices representing the method's syntactic structure. 
    \item \textbf{Java Method Body Extraction}: Utilizes the extracted positional metadata to locate and retrieve the full Java method body from the source file. This ensures that the entire implementation is available for translation, including its local variables, control flow, and API calls. 
    \item \textbf{Specification Retrieval:} Retrieves specifications associated with each method using our Specification Extraction Agent. These specifications describe the method’s purpose, usage scenarios, input/output constraints, and any relevant behavioral expectations, enriching the context available for accurate translation.  
    \item \textbf{Prompt Generation}: Constructs a comprehensive prompt tailored for the LLM. This prompt incorporates the method’s name, source code, AST structure, and its retrieved specification. The combined context enables the LLM to perform semantically aware translation from Java to Swift, accounting for both structural fidelity and intended behavior. 
\end{enumerate}

\textbf{Prompt Design.} We show the prompts used for method-level translation from Android to iOS in Table~\ref{tab:mtprompt}.

\begin{table}[t!]
\caption{Method-level translation prompt}
\label{tab:mtprompt}
\begin{tabularx}{\textwidth}{X}
\hline
You are an expert code translator with deep knowledge of Java, Swift, and Android-to-iOS migration. Your task is to translate ``one Java method'' into idiomatic Swift while preserving its original semantics and behavior.
Please translate the following Java (Android) method titled \{Insert method name here\} from \{Insert file name here\} into Swift.

\textbf{Input:}

\textbf{Method Name:} \{Insert method name here\}

\textbf{Method Code:}
\{Insert method code here\}

\textbf{Abstract Syntax Tree:}
\{Insert Abstract Syntax Tree here\}

\textbf{Retrieved Specification:}
\{Insert Retrieved Specification here\}

\textbf{Output Requirement:}  
\begin{itemize}

 \item  Translate the Java method into Swift, preserving the logic and behavior.

 \item Use idiomatic Swift syntax and appropriate iOS constructs.

 \item Assume the class structure will be assembled separately; focus only on method translation.

 \item Ensure compatibility with Swift’s language features and memory management rules (e.g., optional handling, type inference).

 \item  If the Java method interacts with Android-specific APIs, map them to their Swift/iOS equivalents if available. Otherwise, mark clearly with a comment // TODO: Platform-specific adaptation required.
   
\end{itemize}
\\
\hline
\end{tabularx}
\end{table}

After translating all individual methods within a class, the agent proceeds to translate and reconstruct the complete class structure. This step focuses on integrating the previously translated methods into a coherent class definition and refining the overall translation to ensure structural and semantic integrity. 
The class-level translation process leverages multiple sources of information, including the outputs of method-level translation, the class-level dependency graph, and the AST of the class. By incorporating these elements, the agent ensures that all translated methods are accurately assembled within the correct class context, maintaining inheritance relationships, access modifiers, and member variable declarations. Additionally, the agent addresses class-level constructs such as constructors, field initializations, annotations, and inner classes, adapting them to align with Swift idioms and iOS development conventions. This integration step is critical for preserving the behavior, structure, and design of the original Android codebase while producing a functionally equivalent and idiomatic iOS counterpart. Ultimately, this phase ensures that the translated classes are internally consistent, syntactically valid, and ready for downstream packaging and validation. 
In the class-level translation phase, the agent performs the following key tasks to ensure a coherent transformation from Android (Java) to iOS (Swift): 

\begin{itemize}

\item \textbf{Method Integration:} Gathers all individually translated methods and embeds them into the corresponding Swift class structure, ensuring proper syntax, ordering, and indentation. 

\item \textbf{Class Signature Construction:} Translates the class definition, including access modifiers, class name, inheritance hierarchy, and implemented interfaces, mapping them to appropriate Swift constructs (e.g., class, struct, protocol, or extension). 

\item \textbf{Field and Property Handling:} Converts class fields and member variables into Swift properties, adjusting for type differences, initializers, and visibility rules.

\item \textbf{Constructor Translation:} Identifies and translates constructors (public \textit{ClassName(...)}) into Swift initializers (\textit{init(...)}), ensuring consistency with property initialization and dependency injection. 

\item \textbf{Dependency Resolution:} Ensures that method calls, property references, and imported modules within the class are correctly mapped and updated to reflect changes in translation or naming conventions.

\item \textbf{Contextual Refinement:} Applies final refinements using information of AST and specifications (extracted from Specification Extraction Agent) to align the translated class with its intended functionality, constraints, and usage context.

\end{itemize}


\textbf{Prompt Design.} Table~\ref{tab:ctprompt} shows detailed prompts to ensure contextually informed translations at the class-level.

\begin{table}[t!]
\caption{Class-level translation prompt}
\label{tab:ctprompt}
\begin{tabularx}{\textwidth}{X}
\hline
You are an expert code translator with deep knowledge of Java, Swift, and Android-to-iOS migration. Your task is to translate ``one Java Class'' into idiomatic Swift while preserving its original semantics and behavior. 
Please translate the following Java (Android) class from \{Insert class name here\} into Swift. Please follow idiomatic Swift conventions and ensure the final class is syntactically correct, logically coherent, and suitable for iOS development.

\textbf{Class Content:}
\{Insert class content here\}

\textbf{Translated Methods:}
\{Insert translated methods here\}

\textbf{Abstract Syntax Tree:}
\{Insert AST here\}

\textbf{Dependency:}
\{Insert Dependency here\}

\textbf{Specification:}
\{Insert RAG context here\}

\textbf{Output Requirement:} 

Using the above information, write a complete Swift class that:
\begin{itemize}

 \item Correctly defines the class structure with proper Swift syntax.
 \item Integrates all provided translated methods in the right order.
 \item Defines and initializes properties (formerly Java fields).
 \item Translates constructors to Swift initializers.
 \item Resolves and updates class-level dependencies.
 \item Preserves the semantics of the original Java class.
\end{itemize}
\\ 
\hline
\end{tabularx}
\end{table}

After all individual classes within a component have been translated, the agent proceeds to translate and reconstruct the complete component structure by integrating the translated classes into a coherent, functional unit. This involves analyzing and resolving inter-class dependencies, maintaining logical and architectural consistency, and ensuring that the reconstructed component aligns with iOS development conventions and Swift-specific design patterns. The agent also handles module-level concerns such as file organization, import statements, shared resources, and entry points, effectively bridging class-level translations into a unified, runnable iOS module. 

\textbf{Prompt Design}: The agent uses the prompt shown in Table~\ref{tab:comp} to ensure contextually informed translations at the component level.

\begin{table}[t!]
\caption{Component-level translation prompt}
\label{tab:comp}
\begin{tabularx}{\textwidth}{X}
\hline
You are an expert code translator with deep knowledge of Java, Swift, and Android-to-iOS migration. Your task is to translate ``one Java Component'' into idiomatic Swift while preserving its original semantics and behavior. Please translate the following Java (Android) class from \{Insert component name here\} into Swift.

\textbf{Translated Classes:}
\{Insert translated classes here\}

\textbf{Abstract Syntax Tree:}
\{Insert AST here\}

\textbf{Dependency:}
\{Insert Dependency here\}

\textbf{Specification:}
\{Insert RAG context here\}

\textbf{Output Requirement:} 

Using the above information, write a complete Swift component that:
\begin{itemize}

\item Integrate all translated classes into a coherent iOS component.

\item Adapt Android-specific logic to iOS equivalents.

\item Maintain architectural and behavioral consistency with the original Android component.

\item Resolve any cross-class dependencies or shared state.

\item Ensure that the component conforms to Swift and iOS conventions (e.g., navigation, view lifecycle).

\end{itemize}
Please follow idiomatic Swift conventions and ensure the final component is syntactically correct, logically coherent, and suitable for iOS development.
\\ 
\hline
\end{tabularx}
\end{table}

After all components are translated, the agent further wraps up the translated components along with essential configuration files, resources, and project metadata to assemble a complete, buildable iOS project. This final integration step ensures that all translated Swift classes, modules, and assets are properly organized, linked, and referenced within the Xcode project structure.  

\textbf{Prompt Design}: The prompt used by the agent to wrap the translated components into a complete iOS project is shown in Table~\ref{tab:wrap}. 

\begin{table}[t!]
\caption{Wrapping Up Components Prompt}
\label{tab:wrap}
\begin{tabularx}{\textwidth}{X}
\hline
You are an expert iOS developer. You are given a set of Swift components that were translated from an Android application. Your task is to assemble these components into a complete, runnable iOS project using Xcode conventions. Follow the instructions below:

\textbf{Translated Components:}
\{Insert translated components here\}

\textbf{Dependency:}
\{Insert Dependency here\}

\textbf{Resource:}
\{Insert resource files of the Android app here\}

\textbf{Configuration:}
\{Insert the configuration files of the Android app here\}

\textbf{Specification:}
\{Insert RAG context here\}

\textbf{Output Requirement:} 

Using the above information, write a complete Swift project that:
\begin{itemize}

\item Integrate all translated components into a coherent iOS project.

\item Maintain architectural and behavioral consistency with the original Android project.

\item Resolve any cross-component dependencies.

\item Ensure that the component conforms to Swift and iOS conventions (e.g., navigation, view lifecycle).

\item Include appropriate folders, file organization,  necessary permissions, and any required project configuration files.

\end{itemize}
Please follow idiomatic Swift conventions and ensure the final project is syntactically correct, logically coherent, and suitable for iOS development.
   
\\ 
\hline
\end{tabularx}
\end{table}

\subsection{Code Validation Agent}
\label{sec:3.3}
This agent ensures the correctness, consistency, and integrity of the translated iOS code through a multi-stage validation pipeline. It begins by verifying semantic consistency across the codebase, focusing on the accuracy of inter-file references, class and method dependencies, and the integrity of cross-component interactions by examining the corresponding dependency structures within Android classes and packages. The agent also analyzes whether module imports, platform-specific APIs, and class hierarchies are preserved and correctly mapped in the translated Swift code. This helps mitigate issues such as broken references, misused iOS frameworks, and incorrect mappings of Android constructs to their iOS counterparts, reducing the risk of runtime failures and integration problems. 
Next, the agent performs automated linting and syntax validation using Swift language tooling to detect and correct syntax errors, undeclared identifiers, incorrect type usage, and other language-specific issues. It also checks for common anti-patterns and platform-specific best practices to ensure the code adheres to idiomatic Swift standards. 
Finally, the agent may optionally conduct lightweight build-time verification, such as compiling isolated modules or running static type checks, to further confirm that the translated components are structurally and semantically sound before integration into the final iOS project. The agent performs the following tasks: 
\begin{itemize}
\item \textbf{Dependency Resolution:} Analyzes class-level and component-level dependencies by examining the corresponding dependency structures within Android classes and packages to confirm that all required relationships (inheritance, composition, API usage) are maintained post-translation. Detects and resolves missing or broken links across modules caused by improper translation or reorganization. 

\item \textbf{Syntax Validation}: Uses the \texttt{swiftc} compiler to identify syntax errors in the translated Swift files. Files with syntax errors are flagged, and error details are logged for correction. Employs \texttt{SwiftLint} to enforce coding standards and identify suboptimal patterns or stylistic issues in the translated code.
    
\end{itemize}
The validation process systematically uses identified issues to guide LLMs in refining the generated code. These feedback signals, ranging from syntax errors to semantic inconsistencies and platform-specific violations, are integrated into follow-up prompts, enabling iterative improvements and more reliable translation outputs.

\textbf{Implementation Details.} The Dependency Validation operates as follows: 
\begin{itemize}
\item Inter-File Reference Checking:
The agent scans translated code to ensure that method, class, and module references remain valid across files. It identifies missing definitions or mismatched identifiers introduced during translation and flags them for correction.

\item Dependency Graph Validation:
The agent reconstructs the dependency graph of the translated iOS codebase and compares it to the original Android project’s structure. It verifies that essential dependency relationships (e.g., method calls, class extensions, module imports) are preserved.

\item Platform Compatibility Analysis:
The agent examines dependencies that rely on Android-specific APIs or libraries and checks whether equivalent iOS-native alternatives have been correctly mapped. Flags any unsupported or untranslated platform-specific functionality.



\end{itemize}

 The Syntax Validation operates as follows: 
\begin{itemize}
\item Language-Specific Linting:
The agent will first run automated linters tailored for Swift to detect syntax errors, style violations, and deprecated language constructs in the translated code. This step ensures compliance with standard Swift coding practices.

\item Parsing and Compilation Checks:
The agent then uses static analyzer tools to parse and compile each translated file individually, as well as the complete iOS project, to verify that the syntax is valid and that the code is structurally executable.

\item  Error Localization and Reporting:
The agent further pinpoints the exact location and type of syntax issues (e.g., missing brackets, incorrect type annotations, misplaced modifiers) and logs them for inspection and repair.

\item Feedback Loop for Code Refinement:
Finally, the agent integrates syntax issue reports into a feedback pipeline that guides LLMs to regenerate or patch faulty code segments, ensuring higher-quality final translations.

\end{itemize}

Note that for the syntax validation stages, the identified issues can be repeatedly used to prompt the LLM to refine the generated code. In this study, we performed up to three rounds of iterative refinement for each issue found during the validation process. This iterative prompting allows the LLM to progressively correct semantic inconsistencies, broken references, and syntax errors. The number of repetitions was empirically chosen to balance improvement gains with generation overhead, as we observed diminishing returns in code quality improvements after five iterations. 



\section{Experimental Setup} 
\label{sec:4}

To investigate the effectiveness, challenges, and future potential of LLM-based code translation for mobile applications, we designed experiments aimed at addressing the following research questions:

\subsection{Research Questions}

\begin{enumerate}
\item \textbf{RQ1}: How effective are LLMs 
in translating mobile application code from Android (Java) to iOS (Swift), preserving functionality and dependencies? 

\item \textbf{RQ2}: What are the primary challenges encountered during mobile application code translation using LLMs, and how can these challenges be categorized?

\item \textbf{RQ3}: What opportunities and future directions exist for advancing LLM-based code translation techniques in this domain? \end{enumerate}

\subsection{Subject Repositories} 

To evaluate the effectiveness of our approach to translating Android applications into iOS, we curated a diverse set of open-source, actively maintained Android repositories. These projects were selected to represent a broad spectrum of real-world mobile applications while maintaining practical constraints for manual inspection and analysis. Specifically, we prioritized repositories that span different application domains—such as productivity tools, media players, and utility apps—ensuring our evaluation captures a wide range of architectural patterns, UI interactions, and platform-specific behaviors commonly seen in Android development. 

Importantly, we excluded projects that heavily rely on third-party libraries to minimize external noise and focus our analysis on core application logic and platform-dependent code. This allows us to more accurately assess the capabilities and limitations of LLM-based translation techniques in handling essential mobile development tasks.

Each selected repository strikes a balance between codebase size, functionality diversity, and structural complexity, enabling a systematic and controlled experimental setting. By grounding our evaluation in these representative codebases, we aim to uncover insights into how well large language models preserve application behavior, handle Android-specific APIs, and translate design patterns into their iOS equivalents. Ultimately, this selection strategy helps highlight the practical challenges and opportunities of applying LLMs to real-world, agentic mobile application translation. 
The selected repositories are as follows, and  Table~\ref{tab:subject_repos} summarizes key statistics for each repository. 

\begin{itemize} 
\item \textbf{LeafPic} \cite{leafpic}: An alternative gallery application designed with fluid material design principles. LeafPic is open-source and comprises 132 source files, making it suitable for evaluating translation techniques in a moderately complex application.
\item \textbf{WeatherApp} \cite{weatherapp}: An open-source Android weather application with 39 files. The beginner-friendly app incorporates popular libraries and modern Android design concepts, providing a straightforward but practical test case.
\item \textbf{AndroidTvMovie}~\cite{androidtvmovie}: A lightweight Android TV application with 34 files interacting with The Movie Database API to provide a movie browsing experience. The application leverages the Android TV platform and its specific components.
\item \textbf{MinimalToDo}~\cite{minimaltodo}: A simple and minimalistic to-do list application. It provides a compact codebase for analyzing Android applications' fundamental UI and task management patterns.
\item \textbf{NextCloud}~\cite{nextcloud}: A large-scale cloud storage and collaboration platform. This repository is a more complex test case to evaluate how LLMs handle extensive Android codebases with multiple interacting components.
\end{itemize}

\begin{table}[t!]
\centering
\caption{Overview of Subject Repositories}
\label{tab:subject_repos}
\begin{tabular}{lccc}
\hline
\textbf{Repository} & \textbf{Source Files} & \textbf{LOC} & \textbf{Domain} \\
\hline
\textbf{LeafPic}~\cite{leafpic}        & 132  & 20,177  & Gallery Application \\
\textbf{WeatherApp}~\cite{weatherapp}  & 39   & 3,384   & Weather Forecasting \\
\textbf{AndroidTvMovie}~\cite{androidtvmovie} & 34   & 2,828   & Media Browsing / Android TV \\
\textbf{MinimalToDo}~\cite{minimaltodo} & 27   & 3,079   & To-Do List Application \\
\textbf{NextCloud}~\cite{nextcloud}     & 955  & 149,549 & Cloud Storage / Collaboration \\
\hline
\end{tabular}
\end{table}

\subsection{Subject LLM}
In this work, as a pilot study, we focused on \textbf{GPT-4o}, renowned for its superior performance in code-related tasks~\cite{chen2021evaluating}. GPT-4o has demonstrated state-of-the-art capabilities in understanding, generating, and translating code across various programming languages~\cite{rasheed2025large}. Compared to other models, GPT-4o consistently maintains semantic correctness and produces syntactically valid outputs, outperforming alternatives in tasks involving complex programming constructs~\cite{zhang2025building}. We used GPT-4o with its default parameters and standard token limits to ensure reproducibility and maintain alignment with standard usage practices. While reliance on a fixed context window posed challenges for larger codebases, this was mitigated through modular prompts at the function and class levels.


\subsection{Evaluation Metrics}

To evaluate the performance of our agent-based translation approach from Android to iOS, we employ a set of metrics designed to assess both the completeness and quality of the translated code. These metrics allow us to systematically measure how well the generated iOS code aligns with syntactic and stylistic expectations and to identify common failure modes. The evaluation is based on the following three core metrics:

\begin{itemize}

\item  \textbf{Percentage of Valid Translated Files:} This metric captures the extent to which the translation pipeline can successfully convert Android source files into valid Swift files. A translated file is considered valid if it passes basic compilation or syntax checking without fatal errors and successfully meets our manual verification criteria, which assess whether the core functionalities present in the corresponding Android file have been accurately translated. This metric reflects the overall coverage and effectiveness of the translation process at the file level. A higher percentage of valid files indicates that the pipeline is robust enough to handle structural and contextual elements across different components of the application. 

\item \textbf{Number of Syntax Errors:} Syntax errors are detected using Swift language parsers and compilers. This metric counts the number of distinct syntax violations in the translated code, such as missing brackets, incorrect type declarations, or malformed method signatures. Syntax errors are critical because they prevent the translated code from compiling or running, and they reflect issues in prompt engineering, token limitations, or model understanding of language-level constructs. Fewer syntax errors suggest that the LLM and associated agents are correctly mapping Java syntax to Swift syntax, even in the presence of structural differences between the two languages.

\item \textbf{Number of Lint Issues:} Linting identifies stylistic and semantic issues that, while not necessarily preventing compilation, indicate poor code quality or violations of platform-specific conventions. For iOS code, we use SwiftLint to detect problems such as improper naming conventions, unused variables, improper indentation, and violations of idiomatic Swift practices. This metric serves as a proxy for code maintainability and platform compliance. A low number of lint issues implies that the generated code is not only functional but also adheres to community best practices, making it more readable, reusable, and acceptable to iOS developers.

\end{itemize}

For the metric ``\textbf{Percentage of Valid Translated Files}'', two authors of this paper manually reviewed the translated files to determine their validity, assessing whether the core functionalities were accurately preserved. All disagreements were resolved through discussion with two additional authors. The inter-rater agreement, measured using Cohen’s kappa, was above 0.8 across all files, indicating a high level of consistency and suggesting the reliability and effectiveness of the validation process. 
For the metrics \textbf{Number of Syntax Errors} and \textbf{Number of Lint Issues}, we applied Swift language parsers and SwiftLint to all translated files across the five projects.


It is important to note that, although our agents incorporate mechanisms to ensure the quality of the translated Swift code and successfully generate Swift translations, we cannot build the translated iOS projects successfully without further substantial human efforts. 
To maintain the integrity of the evaluation and reflect the actual performance of the LLM-based translation pipeline, we deliberately refrained from applying manual fixes or interventions to resolve these issues. This decision allows us to assess the out-of-the-box effectiveness of the automated translation approach, without conflating it with human-driven refinements. 
The key challenges associated with the Swift project requirements posed significant obstacles to automated build and runtime validation of the translated code. These challenges include:

\begin{itemize}
   \item \textbf{Dependency Management}: While our agent incorporates a dedicated dependency validation step, many translated files rely on third-party libraries and external frameworks that require additional configuration to integrate correctly with the Swift Package Manager. This dependency complexity significantly complicates the build process. 
   
\item  \textbf{Environment Compatibility}: Fundamental differences between the Android and iOS runtime environments, including reliance on platform-specific APIs and system-level services, made direct compilation of the translated projects infeasible without extensive manual adaptation to bridge these platform disparities. 

\item \textbf{Project Structure Alignment}: Although our agent design accounts for the structural characteristics of the source Android project, the translated Swift code still demands substantial restructuring to conform to Swift's project organization and build conventions. This includes module reorganization, interface adaptation, and configuration of the build system. As a result, the translated projects could not be built successfully without additional, non-trivial engineering effort. 

\end{itemize}

\section{Result Analysis}
\label{sec:5}

\subsection{RQ1: Effectiveness of LLM-Based Code Translation for Mobile Applications}

\noindent \textbf{Approach.} We conducted an extensive evaluation of LLM-based code translation across five subject datasets, as outlined in Table~\ref{tab:subject_repos}. For this evaluation, we employed \textbf{GPT-4o} as the underlying large language model, selected for its strong performance on reasoning and code generation tasks. The translation process followed our proposed multi-agent pipeline in Section~\ref{sec:3}, which decomposes and translates code with greater structural awareness. 
As we claimed in Section~\ref{sec:3.3}, we performed three rounds of interactive refinement for each file for both the dependency resolution and syntax validation. 
A comprehensive summary of the translation performance on each dataset is presented in Table~\ref{tab:translation_results}. 
Note that to illustrate GPT-4o's performance on direct code translation, we present the results both before and after applying the ``Code Validation Agent''. 

\begin{table}[t!]
\centering
\caption{Translation Results for Subject Datasets (Before/After Code Validation Agent)}
\label{tab:translation_results}
\begin{tabular}{lcccc}
\hline
\textbf{Project Name} & \textbf{Total Files} & \textbf{Valid Files} & \textbf{\# Syntax Errors} & \textbf{\# Lint Issues} \\
\hline
LeafPic & 132 & 46.2\% / 78.7\% & 13 / 0 & 68 / 28 \\
WeatherApp & 39 & 69.2\% / 94.9\% & 0 / 0 & 12 / 2 \\
AndroidTvMovie & 34 & 70.6\% / 85.3\% & 0 / 0 & 10 / 5\\
MinimalToDo & 27 & 0 / 44.4\% & 2 / 0 & 25 / 15 \\
NextCloud & 955 & 43.0\% / 68.7\% & 115 / 0 & 429 / 298 \\
\hline \hline
Total & 1,187 & 43.2\% / 70.7\% & 130 / 0 & 544 / 348 \\ \hline
\end{tabular}
\end{table}

\noindent \textbf{Result.} 
As shown in the table, direct code translation using LLMs still suffers from fundamental issues such as syntax errors and linting problems. GPT-4o successfully produced valid translations for a substantial portion of source files across all five projects, with validity rates ranging from 0\% to 70.6\% and an average of 43.2\% before applying the Code Validation Agent. 
The introduction of the Code Validation Agent led to noticeable improvements, increasing the percentage of valid files by 14.7\% to 44.4\%, with an average improvement of 28.6\% across projects. Additionally, GPT-4o's performance varies significantly between projects, likely due to differences in code complexity, architectural patterns, and reliance on platform-specific APIs—all of which impact the difficulty of achieving accurate translations.


Regarding syntax errors and linting issues, we also found considerable variability among the projects. Notably, none of the translated projects exhibited syntax errors after passing through the syntax validation stage, demonstrating the LLM’s capability to generate structurally correct code when supported by automated validation mechanisms.  
However, linting issues were more prevalent than syntax errors. This is likely because linting involves a broader range of concerns, such as coding style, formatting, and best practices, that go beyond basic syntax and are not explicitly encoded in the LLM’s training objectives. Although our validation agent resolves most linting issues across multiple iterations, some persist even after three rounds of refinement. This could be due to limitations in  GPT-4o's fixing heuristics for these lint issues in Swift coding, or new linting violations introduced while addressing prior syntax errors. Two examples are shown below:

\begin{lstlisting}
WeatherApp/HTTPWeatherClient.swift:68:1: warning:
Trailing Whitespace Violation: Lines should not have trailing whitespace (trailing_whitespace)
\end{lstlisting}
\begin{lstlisting}
AndroidTvMovie/GlideBackgroundManager.swift:66:1:
warning: Line Length Violation: Line should be 120 characters or less; 
currently it has 194 characters (line_length)
\end{lstlisting}
The first example is from the translated project \textit{WeatherApp}, where SwiftLint flagged line 66 in the file \textit{HTTPWeatherClient.swift} with a ``Trailing Whitespace Violation''. In this case, the line ends with unnecessary spaces or tabs after the last visible character.  The second example comes from the translated project \textit{AndroidTvMovie}, where SwiftLint identified line 66 in the file \textit{GlideBackgroundManager.swift} as a ``Line Length Violation''. 
The line exceeds the recommended 120-character limit, measuring 194 characters in total. Both kinds of violations don't affect program execution but violate style guidelines. 
These examples illustrate a subtle yet common formatting oversight that may not be detected by LLMs unless explicitly prompted. Future work can focus on enhancing prompt design or applying post-generation formatting tools to help resolve such persistent violations.

Notably, NextCloud, the largest and most complex project in our evaluation, exhibited the highest frequency of both syntax and lint errors before and after the validation stage. This trend aligns with the higher architectural complexity and inter-module dependencies typically found in large-scale codebases, which make automated translation and validation more challenging.





In these cases, repeated patterns of style violations suggest that the translation system does not yet incorporate awareness of idiomatic usage or formatting conventions. These issues, while not preventing execution, hinder code readability and deviate from team or project-level style guidelines. 

Moreover, some lint warnings appear to be false positives, accounting for 18.1\% of unfixable issues (63 out of 348). For example, a trailing whitespace warning may be incorrectly raised on lines that are syntactically altered by the translation agent for padding or alignment, but the manual inspection did not actually detect this issue.

\mybox{\textbf{Answer to RQ1:} LLMs achieved an average valid translation rate of 43.2\% before and 70.7\% after applying the Code Validation Agent across five projects, with performance varying based on code complexity. While syntax errors were resolved, linting issues persisted, particularly in more complex codebases, reflecting the challenges of maintaining style and best practices in automated translation.}

\subsection{RQ2: Challenges and Taxonomy in Mobile Application Code Translation}

\noindent \textbf{Approach.} 
In RQ1, we observed that about half of the translated files were invalid during the initial code translation stage (i.e., before the Code Validation Agent). To systematically investigate the root causes of these translation failures, we conducted a rigorous \textit{manual analysis} of a set of samples that our LLM-based approach failed to translate.   
Note that the subjects of this analysis include both the syntax errors reported by the Swift compiler in RQ1 before and after the Code Validation Agent. 
Among all the samples, we performed random sampling with a 95\% confidence level. As a result, a total of 380 errors and issues, spanning the method, file, and package levels, were analyzed.

Specifically, two of the authors, each with over five years of software development experience, independently examined each unsuccessful translation case to ensure objectivity in classification. During this review, they focused on identifying the \textit{root causes} of translation failures and understanding the scope of the errors. Our analysis also includes an exploration of the impact of errors in the translated Swift code. To systematically study these issues, we categorize them into three levels of granularity, i.e., method-level, where errors affect individual functions or code blocks; file-level, where the issues span across an entire source file and may impact interactions between methods or classes; and package-level, where the errors involve dependencies on external libraries or frameworks, often leading to compatibility or integration problems. 
This hierarchical classification helps us better understand the scope and severity of translation errors and their implications for application correctness and maintainability. For each sample, the analysis includes understanding both the original and translated code segments, comparing them against ground truth implementations, and labeling the failures under relevant categories. When a new root cause category emerged—one not covered by the initial categories—the authors paused their labeling efforts and convened meetings to discuss the emerging pattern. This collaborative approach allowed them to \textit{validate new categories}, adjust the taxonomy structure, and re-label affected cases accordingly. After completing their independent reviews, the authors resolved discrepancies through in-depth discussions, ultimately achieving consensus for each translation failure. Once the labeled category is finalized and no longer updated, the authors assess the scope of influence for each category and comprehensively categorize all independent labels into three levels: method-level, file-level, and package-level. This iterative and collaborative process resulted in a \textit{refined taxonomy} that groups failure scenarios into distinct categories, reflecting various syntactic, semantic, and contextual complexity levels. Below, we present the final taxonomy along with representative examples and insights into how these failures manifest in LLM-based code translation. 



\noindent \textbf{Result.} The taxonomy consists of 10 distinct issue types identified through our manual analysis, spanning three levels of granularity regarding their scope and impact, i.e., method, file, and package. 
Overall,  file-level issues, especially internal reference mismatches (23.95\%), dominate, indicating a challenge in preserving inter-method or inter-class coherence during translation. Also, a significant number of issues (e.g., third-party library mismatch, platform-specific settings/design) suggest that current LLMs lack platform-awareness in translation tasks. Syntax and linting issues remain prevalent underscores the need for robust post-editing or more rigor validation layers. 
The full list of reviewed cases is provided in our replication package. We briefly describe each category and provide an example for demonstration. 

\begin{figure}[H] \centering \includegraphics[width=\linewidth]{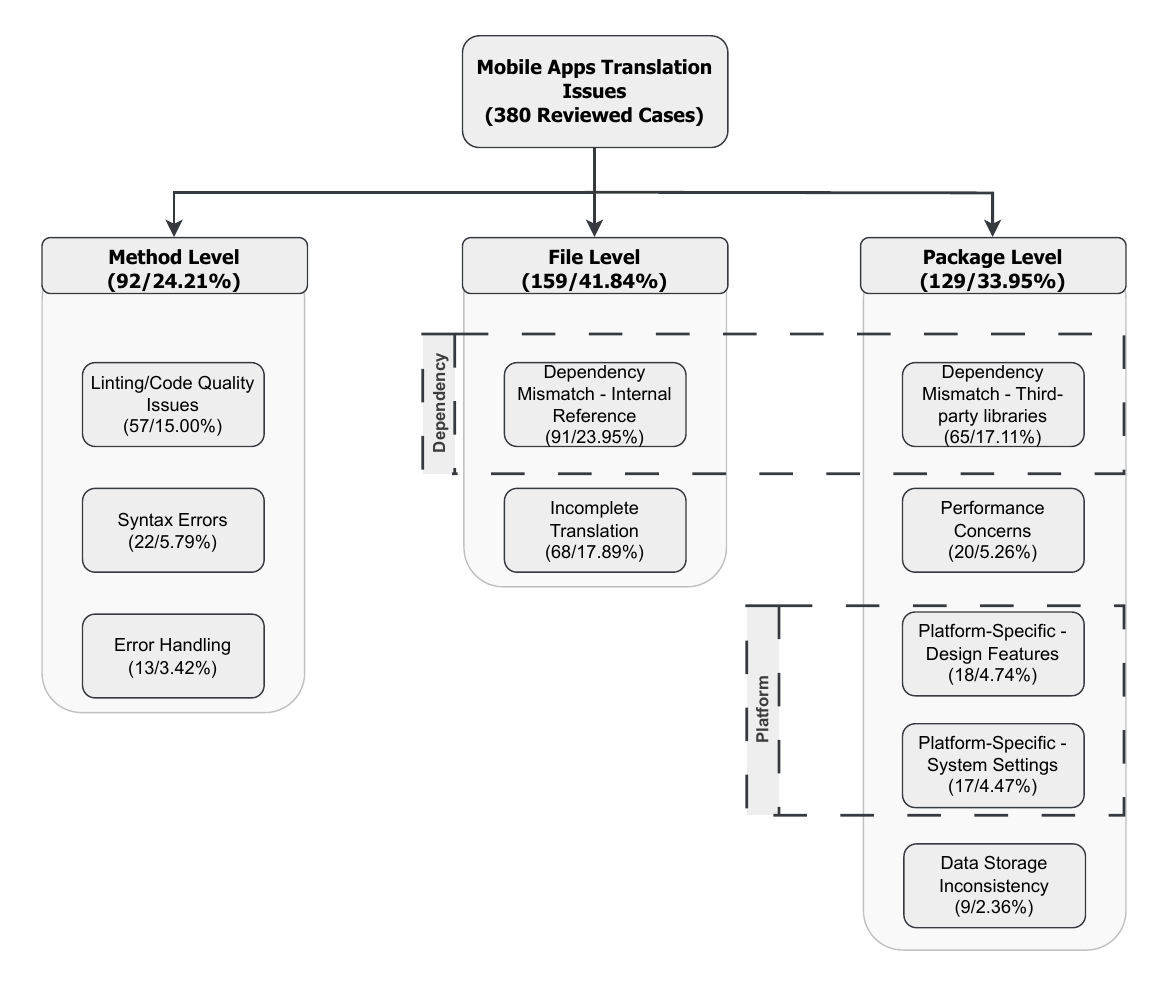} \caption{The categories of translation errors and their distributions} \label{fig:category}\label{Category} \end{figure}

\subsubsection{Method-Level Issues}
Method-level issues impact the correctness and maintainability of individual methods. Over half of the issues at this level are related to linting violations or poor code quality, including inconsistent formatting, naming conventions, and unused variables. In addition to these stylistic concerns, we observed both syntax and semantic errors—such as incorrect API usage, type mismatches, and malformed control structures—that compromise the reliability and readability of the translated methods. Furthermore, the presence of unreachable code and inadequate error-handling mechanisms suggests reduced execution efficiency and robustness, potentially leading to runtime failures or unexpected behaviors in downstream processing.

\noindent \textbf{Linting/Code Quality Issues}:
This is the most frequent method-level issue (13.42\%) in our analysis. These issues refer to the translated code violating coding standards and style guidelines, hindering readability, maintainability, and consistency.

\begin{figure}[H] 
\centering 
  \includegraphics[width=\textwidth]{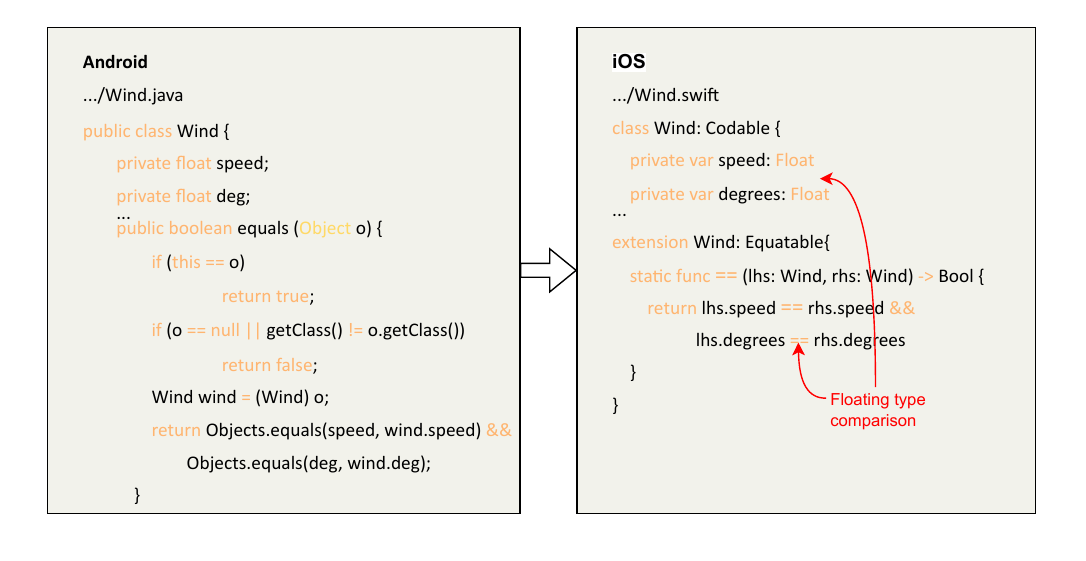}
\caption{An example issue labeled as \textit{Linting/Code Quality}}
\label{fig:sample1}
\end{figure}


Figure~\ref{fig:sample1} illustrates a linting and code quality issue identified in the translation of the WeatherApp project. Specifically, the translated Swift code compares two floating-point values using the \textit{==} operator, while the original Android implementation correctly uses the \textit{equals()} method. Direct comparison of floating-point values with \textit{==} in Swift is generally discouraged due to potential precision and rounding errors inherent in floating-point arithmetic. Such direct comparisons can lead to unexpected behavior and reduced code robustness. To preserve semantic correctness and ensure idiomatic Swift practices, this issue should be addressed by replacing the \textit{==} comparison with an approximate equality check such as \textit{abs(a-b) < epsilon}, where \textit{epsilon} is a small threshold value. 

\begin{figure}[t!] 
    \centering 
    \includegraphics[width=\linewidth]{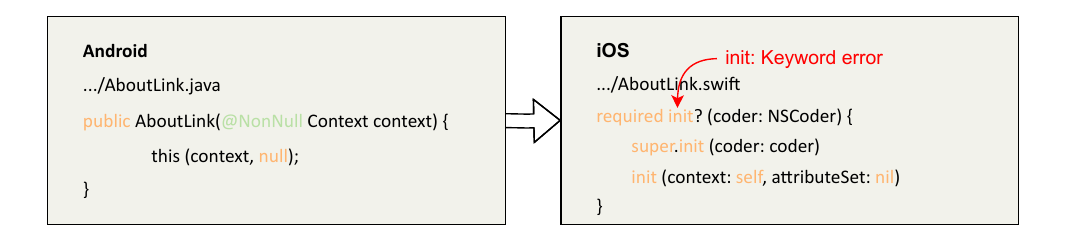}
    \caption{An example issue labeled as \textit{Semantic Error}}
    \label{issue2} 
    \end{figure}

\noindent \textbf{Syntax Errors}: Our analysis also revealed that 5.79\% of the reviewed issues were labeled as Syntax Error, which refers to the incorrect use of the programming language’s grammar or structure, e.g., incorrect use of keywords or invalid identifiers. Figure~\ref{issue2} presents an example of a semantic error in the generated Swift code, where the use of the keyword \textit{init} as an identifier triggered a syntax error. In Swift, \textit{init} is a reserved keyword used for initializers and cannot be repurposed as a variable or function name. Consequently, we categorized this issue under Syntax Errors due to the violation of language-specific naming rules. 

\begin{figure}[t!] \centering \includegraphics[width=\linewidth]{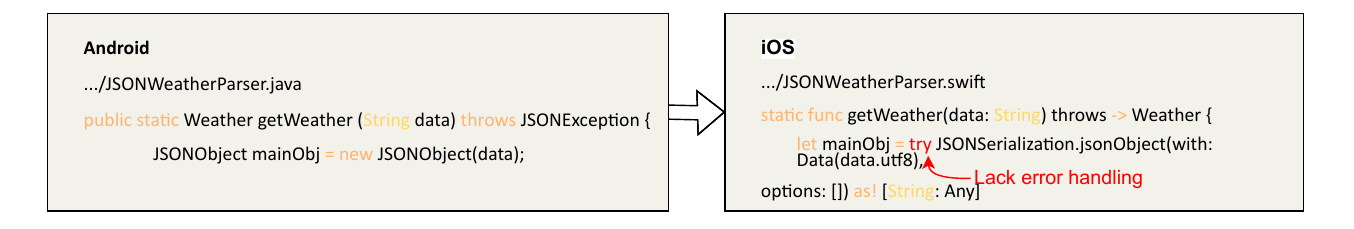}
   
      \caption{An example issue of \textit{Error Handling}}
        \label{issue3}
      \end{figure}

\noindent \textbf{Error Handling}: This category accounts for 3.42\% of all reviewed issues. Errors in this category arise when the translated code handles exceptions, failures, or edge cases improperly, i.e, either by omitting necessary checks, using incorrect constructs, or failing to preserve the original error-handling logic. Figure~\ref{issue3} presents a representative example from the Error Handling category. In this instance, the \textit{getWeather} function in the original code is designed to throw a \textit{JSONException} to handle potential failures arising from the JSONSerialization process. However, the translated Swift code does not include appropriate error-handling constructs. While it uses the try keyword to indicate a potential failure, it lacks the necessary do-catch block or any fallback mechanism to properly catch and respond to the exception. This omission indicates a failure to preserve the original exception-handling semantics and aligns with the characteristics of Error Handling issues.

\subsubsection{File-Level Issues}
File-level issues involve the structure or implementation of a single/multiple classes or a specific file. We identified three types of patterns regarding internal reference, incomplete translation, and framework-API differences.  
It is worth mentioning that the dependency mismatch contains two subcategories; we categorize internal references in the file-level and third-party libraries in the package-level. This classification will be further explained in section \ref{subsec:package-level}.

\begin{itemize}
    \item \textbf{Dependency Mismatch - Internal Reference}:
This category accounts for the largest proportion of all labeled issues, representing 23.95\% of the total. These errors typically arise from missing or unresolved references to custom classes, methods, or constants that are defined elsewhere in the codebase but are not included in the translated or generated code segment. 
In such cases, the model generates code that depends on internal constructs, such as project-specific classes, helper functions, or configuration values, without having access to their definitions or expected context. This leads to compilation errors or runtime failures due to undefined symbols. The issue reflects a broader challenge in LLM-based code generation, i.e., the lack of global project awareness and incomplete modeling of cross-file or cross-package dependencies, which are common in real-world software systems.

\begin{figure}[t!]
    \centering
    \includegraphics[width=\linewidth]{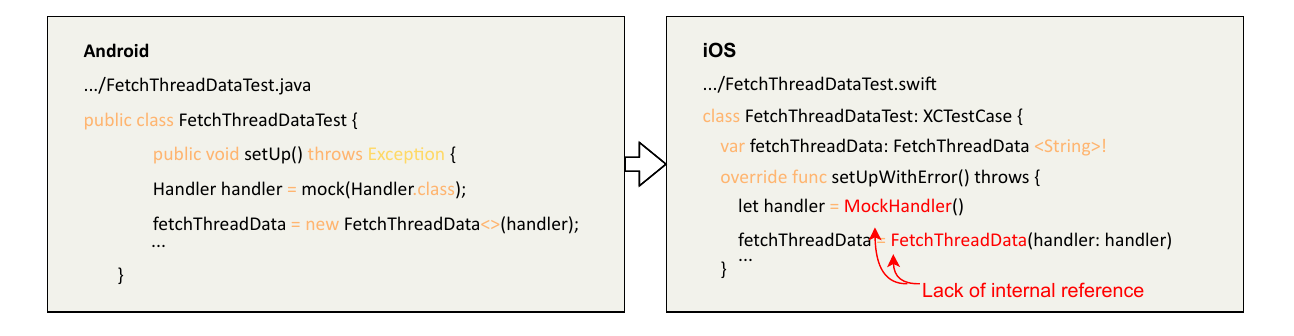}
    \caption{An example issue of \textit{Dependency Mismatch - Internal Reference}}
    \label{fig:6}
\end{figure}


Figure~\ref{fig:6} presents an example from the \textit{WeatherApp} project that illustrates this type of error. In the translated file \textit{DetailFragment}, the method \textit{setUpWithError()} invokes two custom methods, i.e., \textit{FetchThreadData} and \textit{MockHandler}, whose implementations are missing. This results in an internal dependency mismatch, as the generated code references project-specific components that are undefined or inaccessible in the current context.

    \item \textbf{Incomplete Translation}:
    In 16.58\% of the reviewed cases, the translated code appeared to be a direct conversion from an Android-based environment without being fully adapted to the conventions and design patterns of the Swift platform. These cases often reflect inconsistencies in code logic or architectural patterns that are common in Android but misaligned with Swift/iOS development practices. As a result, the translated code may be syntactically correct but functionally inappropriate for the target platform.

    \begin{figure}[t!] \centering \includegraphics[width=\linewidth]{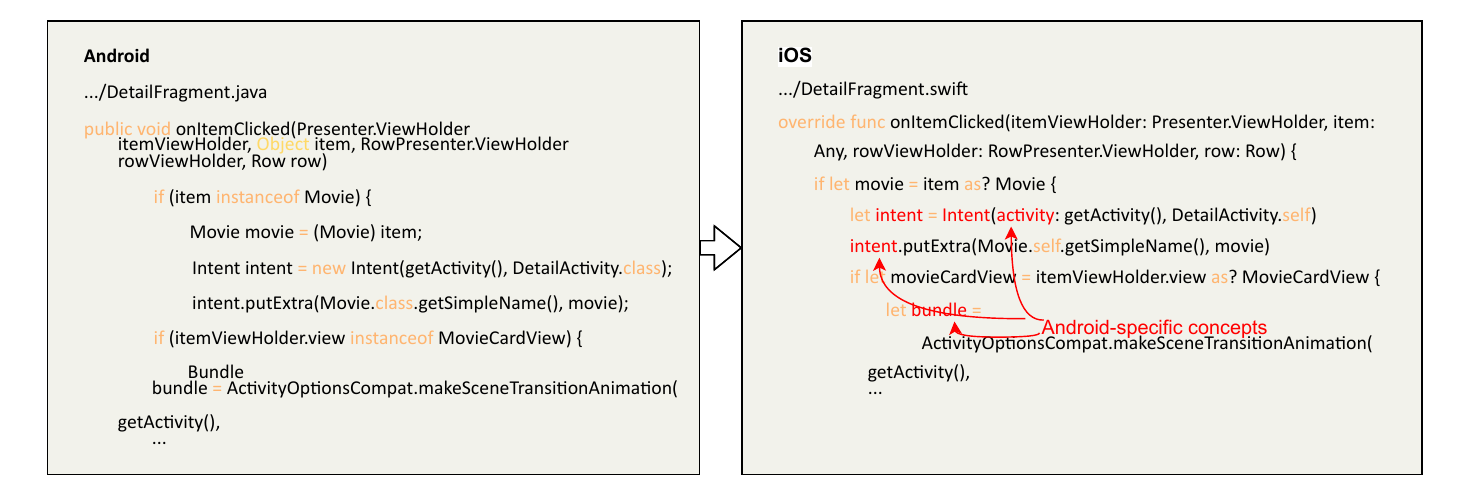}

    \caption{An example issue of \textit{Incomplete Translation}}
        \label{issue7} 
    \end{figure} 

Figure~\ref{issue7} presents an example from the \textit{AndroidTvMovie} project that illustrates an instance of Incomplete Translation. In this case, the translated class includes Android-specific constructs such as Intent, Activity, and Bundle, which are commonly used for navigation and data passing in Android applications. These components should have been replaced with their iOS counterparts, most notably, UIViewController and associated iOS navigation mechanisms, during translation to Swift. However, the translation agent failed to perform this platform-specific adaptation, leaving the Android-specific logic largely intact. This example clearly aligns with our definition of Incomplete Translation, where the output lacks full integration into the conventions and semantics of the target platform.

\end{itemize}

\subsubsection{Package-Level Issues}\label{subsec:package-level}

These issues in this category span multiple files and dependencies, with error patterns at this level potentially impacting the overall project architecture and undermining the robustness and maintainability of the translated application. We identify five categories of such issues, which include dependencies on third-party libraries, platform-specific factors (such as design paradigms and system configurations), data storage mechanisms, and performance-related concerns.

\begin{itemize}

\item \textbf{Dependency Mismatch - Third-party Libraries}: Dependencies on third-party libraries can directly affect the execution of the entire application. We classify mismatches of this kind as package-level issues. In this case, the translated code still relies on third-party libraries that are specific to the Android environment, leading to translation failures. These issues occur when the source code depends on platform-specific frameworks or APIs that lack direct counterparts in the target environment, resulting in semantic mismatches or runtime errors. This underscores the need for platform-aware translation mechanisms that go beyond syntactic conversion and incorporate a deeper understanding of framework-level design patterns and platform compatibility.

 \begin{figure}[t!] \centering \includegraphics[width=\linewidth]{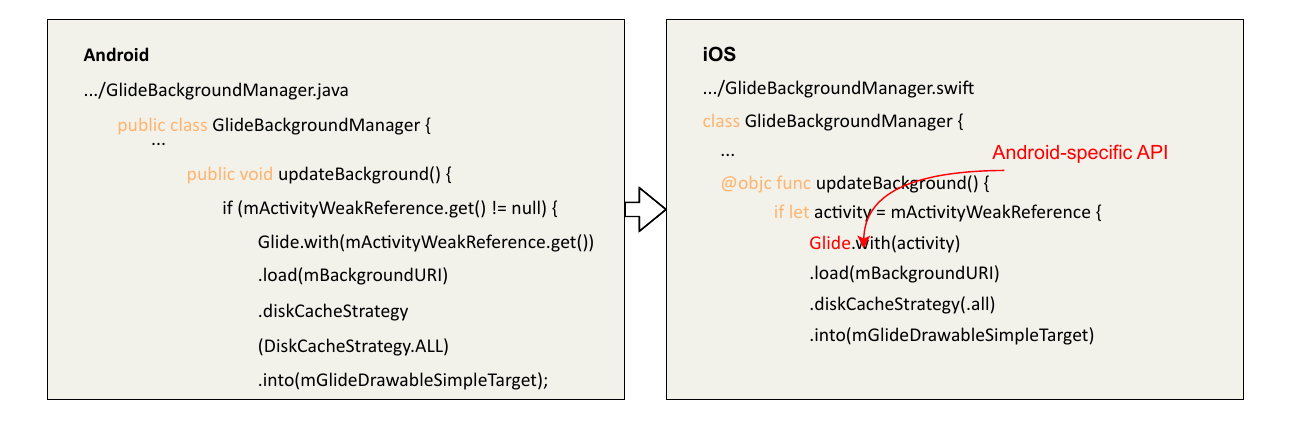}
 
  \caption{An example issue of \textit{Framework/API differences}}
  \label{issue8} 
 \end{figure}


Figure~\ref{issue8} presents an example of this category from the \textit{AndroidTvMovie} project. In this case, the original Android code uses Glide, an Android-specific image loading and processing library, to handle image rendering. However, in the translated Swift code, the LLM incorrectly retains the use of the \textit{Glide.with(...)} method, which is incompatible with the iOS environment. This represents a fundamental mismatch due to the differing UI architectures and framework ecosystems between Android and iOS. Consequently, we categorize this instance under Dependency Mismatch.




    \item \textbf{Performance Concerns}:
    Performance issues arise from suboptimal solutions in the translated code, leading to unnecessary resource consumption and excessive execution in the system. This, in turn, impacts the efficiency and maintainability of the application.

    \begin{figure}[H] \centering \includegraphics[width=\linewidth]{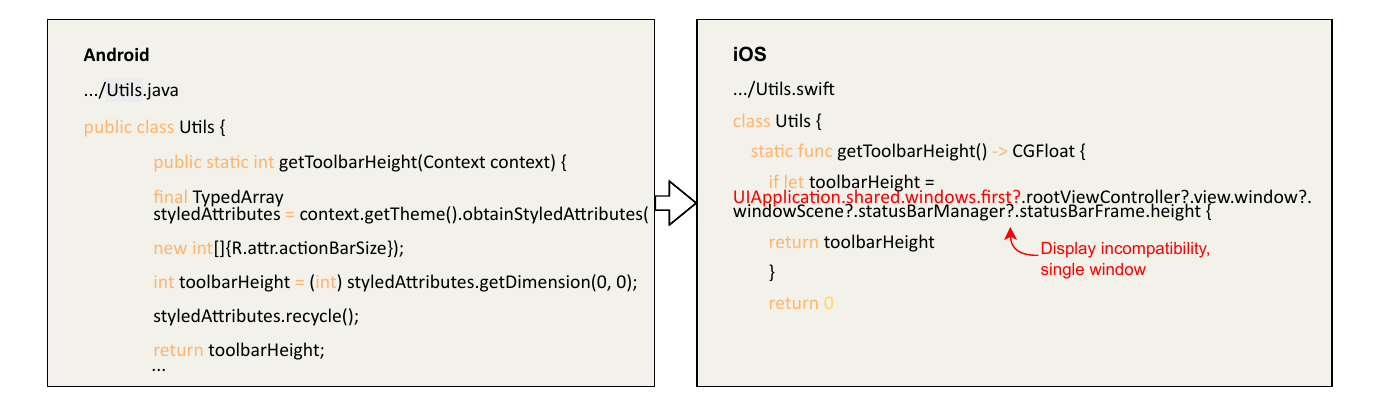}
     \caption{An example issue of \textit{Performance Concerns}}
     \label{issue10}
     \end{figure}

Figure~\ref{issue10} shows an example of Performance Concerns, specifically, in the file \textit{MinimalTodo/Utils.swift}, the code uses \textit{UIApplication.shared.windows.first} to access the window, which is problematic in iOS 13 and later, where apps can have multiple scenes. This approach assumes a single-window setup, which may lead to incorrect behavior in multi-scene environments and reduce maintainability. A more robust solution would involve using \textit{UIApplication.shared.connectedScenes} to correctly identify the active scenes and their associated windows. Due to its inefficiency and potential for compatibility issues with newer iOS versions, this issue is categorized under Performance Concerns. 

\item \textbf{Platform-Specific - Design Features}: Issues in this category typically stem from the improper handling of platform-specific coding patterns and architectural principles, particularly those involving UI components and lifecycle management. 

    \begin{figure}[t!] \centering \includegraphics[width=\linewidth]{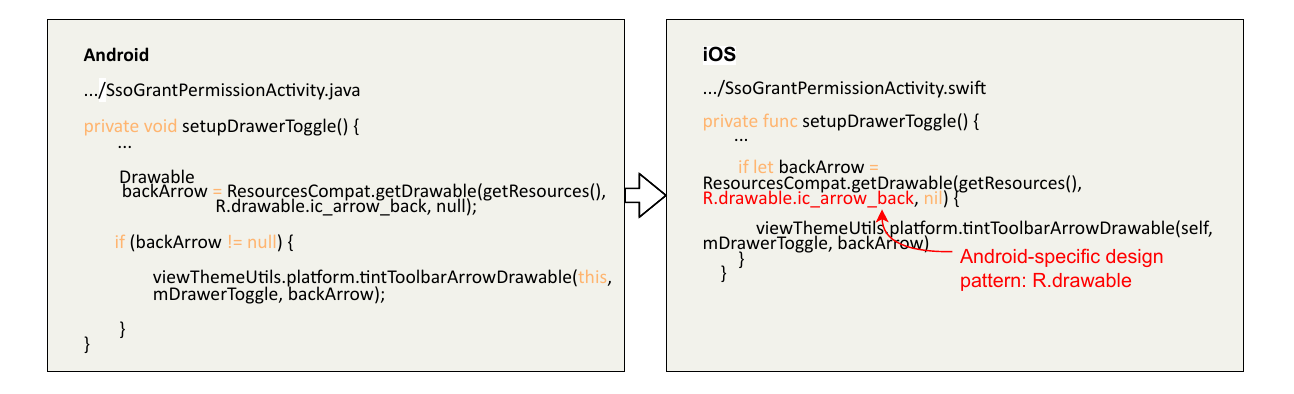}
     \caption{An example issue of \textit{Platform-Specific-Design-Features}}
     \label{issue11} 
     \end{figure}

%

Figure~\ref{issue11} shows an example of this category.  In this case, the translated code erroneously retains Android-specific resource referencing patterns, such as \textit{R.drawable.ic\_arrow\_back}. 
However, iOS follows a fundamentally different approach to asset management, i.e., visual and media resources are organized using Asset Catalogs (\textit{.xcassets}) and accessed programmatically through APIs like \textit{UIImage(named:)}. This direct reuse of Android-style resource access results in incompatibility and reflects a failure to adapt to the design conventions and tooling of the iOS platform. As such, we classify this issue as a Platform-Specific Design Feature problem, where design-related resources are not appropriately translated or mapped to their iOS equivalents, leading to potential UI inconsistencies or runtime errors.


\item \textbf{Platform-Specific-System-Settings}: 
Another category of platform-specific issues arises from differences in essential system features between Android and iOS. These include how each platform handles permissions, data synchronization, background execution, notifications, and other core system behaviors. For instance, Android uses a declarative permission model via the AndroidManifest.xml, whereas iOS requires explicit handling of permissions (e.g., for camera, location, notifications) using runtime authorization requests. Similarly, background tasks in Android are often implemented using Services, WorkManager, or AlarmManager, while iOS relies on Background Tasks, NSURLSession, and strict OS-level constraints. 
   Issues that are classified under Platform-Specific System Features often require platform-aware adaptation rather than one-to-one translation.

    \begin{figure}[t!] \centering \includegraphics[width=\linewidth]{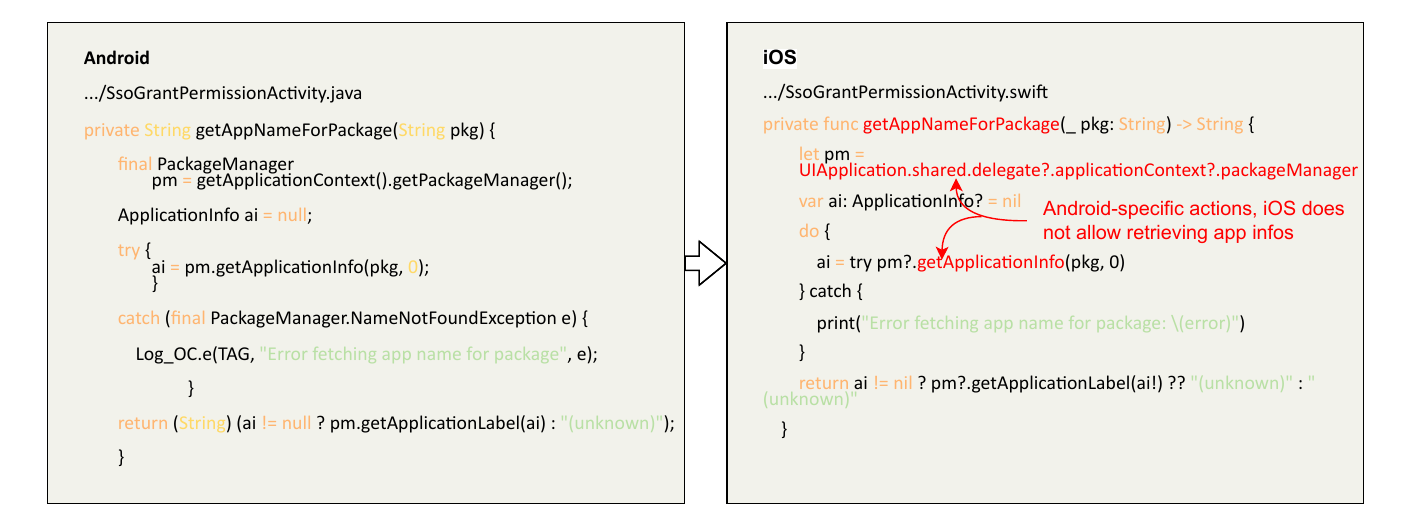}
     \caption{A example issue under Platform-Specific-System-Settings}\label{issue12} 
    \end{figure}



Figure~\ref{issue12} illustrates an example of this category. In this case, the synchronization logic conflicts with iOS system constraints, preventing the retrieval of application information. We classify this issue under \textit{Platform-Specific – System Settings}, as it highlights incompatibilities in handling core system features. The method \textit{getAppNameForPackage} relies on Android-specific APIs to access application metadata, which are not applicable in the iOS environment. This functionality must be reimplemented using iOS-native approaches, such as leveraging Bundle or UIApplication. 

    
\item \textbf{Data Storage Inconsistency}: Android and iOS adopt distinct approaches to data storage and management. Android commonly uses SQLite for structured local data storage, whereas iOS typically favors Core Data, a higher-level framework that integrates with the iOS ecosystem. Directly translating Android's SQLite-based classes and logic to iOS without adapting to Core Data or other native storage mechanisms leads to incompatibilities and potential data management issues. To ensure functional correctness and maintainability, the data layer must be redesigned to align with platform-specific storage paradigms. 

    \begin{figure}[t!] \centering \includegraphics[width=\linewidth]{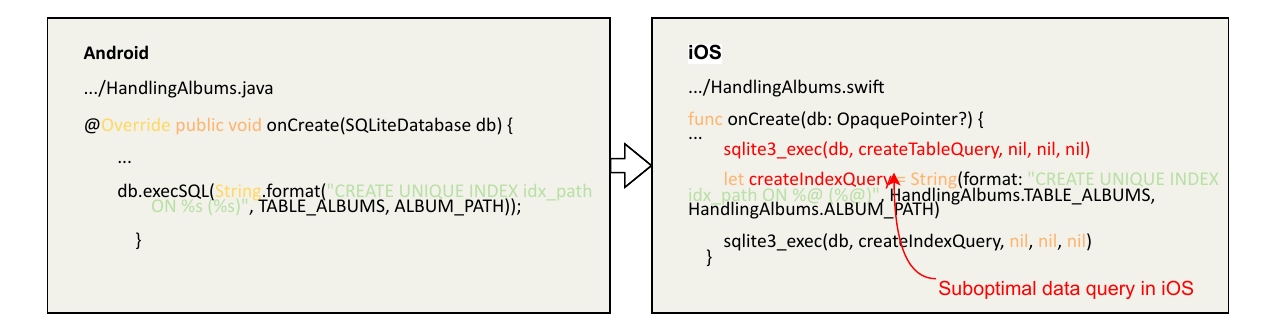}
 
      \caption{An example issue under \textit{Data Storage Inconsistency}}
         \label{issue13} 
    \end{figure}


Figure~\ref{issue13} presents an example of this category. The class interacts with SQLite directly, which is functionally acceptable; however, it would benefit from leveraging a higher-level abstraction such as Core Data or a Swift-friendly third-party library like GRDB. These alternatives offer better integration with the iOS ecosystem and simplify database operations. This example highlights a case of direct translation from Android’s SQLite-based data storage approach without adapting to iOS's native data management paradigms. Continuing to use low-level SQLite in Swift may introduce unnecessary complexity, particularly in handling concurrency, data persistence, and overall maintainability. 

\end{itemize}

\mybox{\textbf{Answer to RQ2:} LLM-based mobile app translation faces challenges across method, file, and package levels. Common issues include syntax errors, incomplete logic, and platform-specific mismatches. Our taxonomy reveals the need for platform-aware translation and rigor validation tools to ensure correctness, maintainability, and compatibility when adapting Android apps to iOS environments.}

\subsection{RQ3: Opportunities and Future Directions in LLM-Based Code Translation}

LLMs have demonstrated remarkable capabilities in translating source code across programming languages and platforms. However, achieving robust, reliable, and maintainable code translation, particularly between complex ecosystems such as Android and iOS, remains a significant challenge. We outline several promising opportunities and future research directions to improve the quality, trustworthiness, and practicality of LLM-based code translation.

\noindent \textbf{ Fine-Tuning for Platform-Specific Semantics:} Fine-tuning LLMs for platform-specific semantics is a crucial step toward improving the quality and reliability of code translation between mobile ecosystems like Android and iOS. While general-purpose LLMs such as GPT-4 are proficient in generic programming tasks, they often lack the nuanced understanding of platform-specific constructs required for accurate cross-platform translation. Android and iOS differ significantly in their architectural patterns, lifecycle management, UI frameworks, permission handling, and asynchronous programming models. For example, translating an Android Activity or Fragment to an iOS UIViewController requires not only matching method names and data flow but also correctly mapping the lifecycle events and navigation paradigms. Likewise, Android’s use of LiveData, Coroutines, and XML-based layouts contrasts sharply with iOS’s Combine framework and SwiftUI's declarative UI style. These differences create significant challenges for LLMs trained only on general code corpora, often resulting in API hallucinations, missing behavioral logic, or syntactically valid but semantically incorrect translations. To address these challenges, fine-tuning should involve curated, high-quality parallel corpora of Android and iOS code pairs, ideally aligned at the method or feature level, covering UI components, system services, and common app functionalities.
Additionally, incorporating static and dynamic analysis data, such as call graphs, control-flow information, and runtime traces, can provide a semantic backbone to help preserve application behavior across platforms. Fine-tuned models should be evaluated not only on syntactic correctness but also on lifecycle fidelity, API equivalence, idiomatic usage, and behavioral correctness using test-based or differential validation techniques. For example, a translation should preserve how and when runtime permissions are requested, how background tasks are handled, and how UI events are propagated.

\noindent \textbf{Human-in-the-Loop Feedback and Corrections:} We also observed that LLMs often fall short in preserving nuanced design decisions, handling non-obvious performance trade-offs, or adhering to implicit architectural conventions. These shortcomings are especially evident in projects where business logic, performance optimization, or legacy constraints are not fully documented in the source code. Human developers bring contextual understanding and domain expertise that current LLMs cannot replicate. Therefore, incorporating human-in-the-loop (HITL) workflows offers a powerful way to improve translation accuracy, maintain semantic integrity, and iteratively refine model behavior. Specifically, future translation agents should provide developer-in-the-loop support where LLM suggestions are presented with the ability to approve, reject, or modify components of the translated code. For example, developers could hover over a generated function and ask, ``Why did you choose NavigationView instead of NavigationStack?'', prompting the model to explain and potentially revise the output. In addition, rather than accepting full translations, developers should be able to give localized feedback (e.g., ``this function incorrectly uses a synchronous API'' ). This feedback can be used in future iterations via few-shot prompting or embedded into active learning loops to fine-tune the model over time.

\noindent \textbf{Third-Party Dependency Synchronization:} 
A significant challenge in cross-platform mobile translation lies in the inconsistent or missing mapping of third-party libraries and SDKs between source and target platforms. In our analysis, we observed frequent loss or misalignment of third-party dependencies during the transition from Android to iOS, which can lead to functional degradation, runtime errors, or incompatibility with project-specific workflows. Current LLM agents lack robust mechanisms to recognize, interpret, and adapt external library usage across platform boundaries. To address this, future research should focus on developing automated dependency synchronization strategies that enable LLMs to identify third-party libraries in the source platform, determine their functional role (e.g., networking, image loading, analytics, database access), and suggest semantically equivalent alternatives available in the target platform ecosystem. For example, a model should be able to translate the use of Android's Retrofit or OkHttp into appropriate iOS networking libraries such as Alamofire or URLSession, or suggest re-implementation using native APIs when no direct equivalent exists. Additionally, integration with build systems (e.g., Gradle, Swift Package Manager) is also essential for automatically resolving and injecting translated dependencies into the target environment. By enabling context-aware, semantically guided synchronization of third-party dependencies, LLM-based translation agents can produce more complete, functional, and production-ready cross-platform code with reduced developer intervention. 

\noindent \textbf{Platform-Specific Adaptation:} 
Our analysis also revealed key categories of platform-specific challenges, e.g., differences in system settings and design paradigms, that often lead to incorrect or suboptimal translations between Android and iOS. These issues highlight the limitations of current LLMs in adapting to the distinct architectural, behavioral, and policy-driven characteristics of each mobile platform. For example, background execution policies, notification handling mechanisms, and energy management constraints differ substantially between Android and iOS, often requiring platform-aware handling in translated code. Similarly, permission management workflows, such as runtime permission requests, implicit vs. explicit permission declarations, and user notification protocols, follow different models that must be correctly reflected in the target platform to preserve functional equivalence and security compliance. Furthermore, platform-specific UI design conventions, accessibility practices, and navigation hierarchies introduce additional complexity when translating across ecosystems. In future work, practitioners should focus on enhancing LLMs with explicit knowledge of these platform-specific behaviors, potentially through fine-tuning on platform-aligned datasets, prompt engineering guided by platform constraints, or the integration of symbolic rules derived from SDK documentation. 

\noindent \textbf{Assessing Post-Translation Effort and Practical Utility:} Future work should also focus on assessing the manual effort required to make automatically translated code fully functional, structurally sound, and semantically correct. While LLMs can produce promising initial translations, developers often need to invest significant effort to resolve structural inconsistencies, adapt platform-specific features, and fix semantic mismatches. Quantifying this post-editing effort would provide a more realistic evaluation of the practical utility of LLM-based translation systems.



\mybox{\textbf{Answer to RQ3:} Future work of LLM-based mobile translation should focus on fine-tuning with aligned datasets, integrating human-in-the-loop feedback, and enabling intelligent adaptation to platform-specific APIs, behaviors, and architectural patterns to improve translation reliability and practicality.}
\section{Threats to Validity}
\label{sec:6}

Our approach to LLM-based mobile application code translation is subject to certain threats to Validity. Below, we outline the key threats and briefly discuss potential mitigation strategies.


\textbf{Internal Validity.} A potential internal threat lies in relying on a single LLM (i.e., GPT-4o) to power the proposed code translation agents. Future work should investigate the impact of using different LLMs to assess the generalizability and robustness of the approach. Additionally, manual validation may introduce subjectivity. To mitigate this, we had two independent participants perform the labeling, cross-referencing their results with documentation and test cases to ensure consistency and reduce bias. 

\textbf{External Validity.} One significant external threat is the generalizability of our findings. While we focused on Android repositories with different sizes for the translation task, the results may not extend to complex projects or other language pairs. Future work will replicate the approach across diverse datasets and programming languages to enhance generalizability.

\textbf{Construct Validity.} The validity of our constructs depends on the quality of tools and techniques used in the pipeline. By employing well-established tools, such as Tree-sitter for AST parsing and reliable vector databases for RAG, we aim to ensure that the methodology is robust and the results are credible.

\section{Conclusion}
\label{sec:7}

In this paper, we take the initiative to investigate the effectiveness of an LLM-based agentic approach for mobile application translation, specifically focusing on migrating Android applications to iOS. By evaluating five Android projects of varying sizes and complexities, we identify key challenges in automated translation. Through manual analysis, we further categorize 10 types of failures with impact scope at the function, file, and package levels, providing insights into common pitfalls and opportunities for improvement. Our findings underscore the potential of LLMs in automating cross-platform migration while highlighting the need for enhanced accuracy, robustness, and platform-specific adaptation. This work lays a foundation for future research aimed at improving the reliability and practical adoption of LLM-driven mobile application translation. 


\bibliographystyle{ACM-Reference-Format}
\bibliography{database}


\end{document}